\documentclass[apjl,iop]{emulateapj}
\pdfoutput=1
\usepackage[fleqn]{amsmath}
\usepackage{graphicx}
\usepackage[varg]{txfonts}
\usepackage{natbib}
\usepackage{float}
\usepackage[usenames,dvipsnames]{xcolor}
\graphicspath{{figures/}}

\renewcommand{\vec}[1]{\boldsymbol{#1}}

\begin{document}

\title{Hydrodynamic moving-mesh simulations of the common envelope phase in
  binary stellar systems}
\shorttitle{Hydrodynamic simulations of the common envelope phase}
\shortauthors{S.~T.~Ohlmann et al.}

\author{Sebastian~T.~Ohlmann$^{1,2}$,
       Friedrich~K.~R\"{o}pke$^{1,3}$,
        R\"{u}diger~Pakmor$^{1}$, and
       Volker~Springel$^{1,4}$
       }
\altaffiltext1{%
    Heidelberger Institut f\"{u}r Theoretische Studien,
    Schloss-Wolfsbrunnenweg 35, 69118 Heidelberg, Germany
}
\altaffiltext2{%
    Institut f\"{u}r Theoretische Physik und Astrophysik,
    Universit\"{a}t W\"{u}rzburg, Emil-Fischer-Str. 31, 
    97074 W\"{u}rzburg, Germany}
\altaffiltext3{%
  Zentrum f\"ur Astronomie der Universit\"at Heidelberg,
  Institut f\"ur Theoretische Astrophysik, Philosophenweg 12,
  69120 Heidelberg, Germany
}
\altaffiltext4{%
  Zentrum f\"ur Astronomie der Universit\"at Heidelberg,
  Astronomisches Recheninstitut, M\"{o}nchhofstr. 12-14, 69120
  Heidelberg, Germany
}

\begin{abstract}
The common envelope (CE) phase is an important stage in binary stellar
evolution. It is needed to explain many close binary stellar systems, such as
cataclysmic variables, Type Ia supernova progenitors, or X-ray binaries. To
form the resulting close binary, the initial orbit has to shrink, thereby
transferring energy to the primary giant's envelope that is hence ejected.  
The details of this interaction, however, are still not understood.  
Here, we present new hydrodynamic simulations of the dynamical
spiral-in forming a CE system.
We apply the moving-mesh code \textsc{arepo} to follow the interaction of a
$1M_\odot$ compact star with a $2M_\odot$ red giant possessing a
$0.4M_\odot$ core.
The nearly Lagrangian scheme combines advantages of smoothed particle
hydrodynamics and traditional grid-based hydrodynamic codes and allows us to
capture also small flow features at high spatial resolution. 
Our simulations reproduce the initial transfer of energy and angular
momentum from the binary core to the envelope by spiral shocks seen in previous
studies, but after about 20 orbits a new phenomenon is observed. Large-scale flow
instabilities are triggered by shear flows between adjacent shock layers. These
indicate the onset of turbulent convection in the common envelope, thus altering the
transport of energy on longer time scales.
At the end of our simulation, only 8\% of the envelope mass is ejected.
The failure to unbind the envelope completely may be caused by processes on thermal
time scales or unresolved microphysics.
\end{abstract}


  \maketitle

\section{Introduction}
\label{sec:introduction}

Many relevant astrophysical processes involve a compact star in a close binary
system, e.g., cataclysmic variables, Type Ia supernova progenitors, X-ray binaries, or
neutron star mergers. In the evolution of these systems, a giant in a wide orbit
ejects its envelope while the orbit shrinks due to interaction with the binary
companion in a common envelope (CE) event. Unstable mass transfer initiates a CE
phase, followed by a rapid spiral-in and possibly further evolution on thermal
time scales \citep[for a recent review, see][]{ivanova2013a}. The first ideas
were developed by \citet{paczynski1976a}, but the problem is still far from
being understood today. Hence, binary population synthesis codes model CE phases
using parametrized prescriptions where the uncertainty of the outcome is
dominated by the parametrization of these phases (compare, e.g., the study by
\citealp{meng2011a} for Type Ia supernova progenitors). As the CE phase does not
possess intrinsic symmetries, hydrodynamical simulations in three dimensions are
required to model the physical processes. Some processes may take place on a
thermal time scale where hydrodynamic simulations are not feasible today and
should be complemented by one-dimensional simulations including more physics,
e.g., energy transport \citep[see the discussion in][]{ivanova2013a}.

Recent hydrodynamic simulations include the adaptive mesh refinement (AMR)
simulations by \citet{ricker2008a,ricker2012a} and the smoothed particle
hydrodynamics (SPH) and unigrid simulations by \citet{passy2012a}.  In these
calculations, the dynamical spiral-in lasts between 10 and 100~d, during which spiral
shocks redistribute angular momentum in the envelope. At the end, only a small
fraction of the envelope becomes unbound and the final separation seems to be larger
than in observed post-CE systems (compare Fig.~17 of \citealp{passy2012a}).
The failure to eject the envelope in current simulations
may possibly be overcome by including recombination energy 
\citep[see SPH simulations by][]{nandez2015a}.

To improve the understanding of hydrodynamical processes during the spiral-in
phase of a CE event, we have run a high-resolution simulation using the
moving-mesh code \textsc{arepo} \citep{springel2010a}. This code solves the
Euler equations on a moving computational grid with adaptive resolution, thus
combining the advantages of traditional SPH and AMR codes, e.g., conservation of
angular momentum and total energy and resolution of low-mass flows and
small-scale flow features. Here, we show that the
\textsc{arepo} code, developed originally for cosmological simulations
\citep[e.g.,][]{vogelsberger2014b,marinacci2014a}, 
allows us to resolve the hydrodynamical structure of the CE phase in
unprecedented detail.
In our simulation, shear flows lead to large-scale Kelvin-Helmholtz
instabilities that dominate the flow structure. These
instabilities have not been observed previously in simulations and may mark the
onset of convection, thus changing the transport of energy in the envelope on a
thermal time scale.

\section{Numerical Methods and Setup}
\label{sec:numericalmethods}

We simulate the dynamical spiral-in of the CE phase with the finite volume
hydrodynamics code \textsc{arepo} \citep{springel2010a}. \textsc{arepo} solves
the Euler equations on a moving, unstructured Voronoi mesh using an HLLC-type
approximate Riemann solver. Self-gravity is included with a tree-based
algorithm. We employ an improved gradient estimate and time integration scheme
\citep{pakmor2016a} yielding second order convergence also for general mesh
motions. Individual and adaptive time stepping is used, which boosts the computational
efficiency due to the multi-scale nature of the system: not all cells are
evolved on the shortest time step, but only the cells requiring the smallest
time steps. 
Periodic boundary conditions were chosen for the hydrodynamics solver with the box
size large enough ($3.3\times 10^{14}$~cm), such that no mass flows
over the boundary for a sufficiently long time. The simulation was stopped when
the first outflow reaches the boundary.

The binary system was set up by placing a red giant (RG) model on the grid with
the core replaced by a gravitation-only particle and then adding another
gravitation-only particle to model a non-resolved, compact
companion star, e.g., a main sequence star or a white dwarf. Thus, the simulation
consists of cells representing the gas of the envelope and the background (``gas
cells'') and gravitation-only particles representing the RG core and the
companion (``core cells'').   

For preparing the single-star initial conditions, we followed Ohlmann et al.
(2015, in prep.). The RG model was created using the stellar evolution code MESA
\citep{paxton2011a,paxton2013a} with a zero-age main sequence mass of
$2M_\odot$. To limit the range in time scales, we replaced the core of the RG by
a particle interacting only gravitationally. We mapped the resulting model to a
grid with mass-adaptive radial shells using a HEALPix distribution
\citep{gorski2005a} on each shell.  The gravitational force of
the particle was smoothed at a length of $h=7.3\times 10^{10}$~cm ($\sim 1.0
R_\odot$) according to the spline function given in \citet{springel2010a}. 
This enables us to reach a stable configuration around the particle since the
pressure gradients can be resolved sufficiently to counteract the gravitational
force of the particle.
We treat the gas as an ideal gas with an adiabatic index of
5/3 which is different from the MESA equation of state. 
However, since we are interested in the envelope where the departure from ideal
gas behaviour is small, this approach still allows for a reasonable representation
of the star calculated with MESA in \textsc{arepo}. 
The mechanical structure of the star is well reproduced in the envelope
(differences in density, pressure, and sound speed are less than 5\%).
Only in the internal energy we observe larger deviations, because we neglect
the ionization state of the gas as a first approximation.
The RG atmosphere was then relaxed by employing
an additional damping term to reduce spurious velocities for several dynamical
time scales (for details, see Ohlmann et al., 2015, in prep.). This resulted in
a stable profile with a core mass of $0.38 M_\odot$ and an
envelope mass of $1.60 M_\odot$ (total: $1.98 M_\odot$). 
The Mach numbers in the envelope reach up to 0.1 after the relaxation
procedure, similar to what would be expected from the initial MESA model,
although we are not able to properly resolve the convection in the envelope.
The density and pressure profiles are stable for several dynamical time scales
after relaxation without applying any damping.
The total number of
cells was about $1.8\times10^{6}$ with a mean cell mass of $8.7 \times
10^{-7} M_\odot$ at the beginning and $2.7\times10^{6}$ with
$5.8\times10^{-7}M_\odot$ at the end of the simulation due to mesh refinement.
The refinement criterion was set to a target cell mass of
$8.4\times10^{-7}M_\odot$. Additionally, in a sphere of five softening lengths
of the gravitation-only particle, the maximum cell radius was bound to a tenth
of the softening length.  The smallest cells near the RG core have a radius of
about $4.6\times10^9$~cm ($0.07 R_\odot$) at the beginning and about
$8\times10^8$~cm ($0.01 R_\odot$) at the end of the simulation.
This allows us to study small-scale flow features in detail. 
In the direct vicinity of the RG core, the hydrodynamical flow is only
resolved outside of a sphere of radius of the softening length. Nevertheless, we
find that a resolution of about 10 cells per softening length is required to
ensure energy conservation during the in-spiral.
The spatial resolution in recent hydrodynamics simulations of CE phases was
lower: the AMR simulations of \citet{ricker2008a,ricker2012a} have $0.3R_\odot$
cells for a $32R_\odot$ RG, the SPH simulations of \citet{passy2012a} have
$0.1R_\odot$ smoothing lengths, and the grid simulations of \citet{passy2012a}
$1.7R_\odot$ and $3.4R_\odot$ cells for a $83R_\odot$ RG.

The companion was placed on the surface of the RG at the same $y$ and $z$
coordinates as the RG core and at a distance of $49 R_\odot$ in the $x$ direction.
The mass of the secondary was chosen to be $0.99 M_\odot$, half of the primary
mass.  The velocities were initialized to a rigid rotation around the center of
mass with the Keplerian rotation period of 23~d.  Moreover, the envelope was
assumed to be in 95\% co-rotation, similar to the simulations of
\citet{ricker2008a,ricker2012a}. 
More realistic initial conditions would start at the point of
Roche lobe overflow to take into account the transfer of energy and angular
momentum from the orbit to the envelope. Moreover, the hydrostatic equilibrium
in the outer layers is less distorted compared to placing the companion at the
surface.  However, since the time scales of orbital decay are very long when the
first mass transfer starts, it is at the moment computationally infeasible for
us to start the simulation at the Roche lobe distance.

The RG atmosphere and the companion were placed in a large box with a side
length of $3.3\times 10^{14}$~cm and a low background density\footnote{The
density to which we resolve the initial RG model is 0.002~g~cm$^{-3}$.} of
$10^{-16}$~g~cm$^{-3}$. 
To resolve the gravitational interaction between the cores, the
softening lengths of the cores are required to be at most a fifth of the distance
between the cores.
The simulation was stopped after about $120$~d,
when the first low-density tidal arm reached the boundary. Since the box was
chosen very large, no mass is lost during the simulation.  Angular momentum was
conserved to high accuracy during the run with an error below 1\%.
  
\section{Results and Discussion}
\label{sec:results}

\begin{figure}[btp]
  \centering
  \includegraphics[width=\columnwidth]{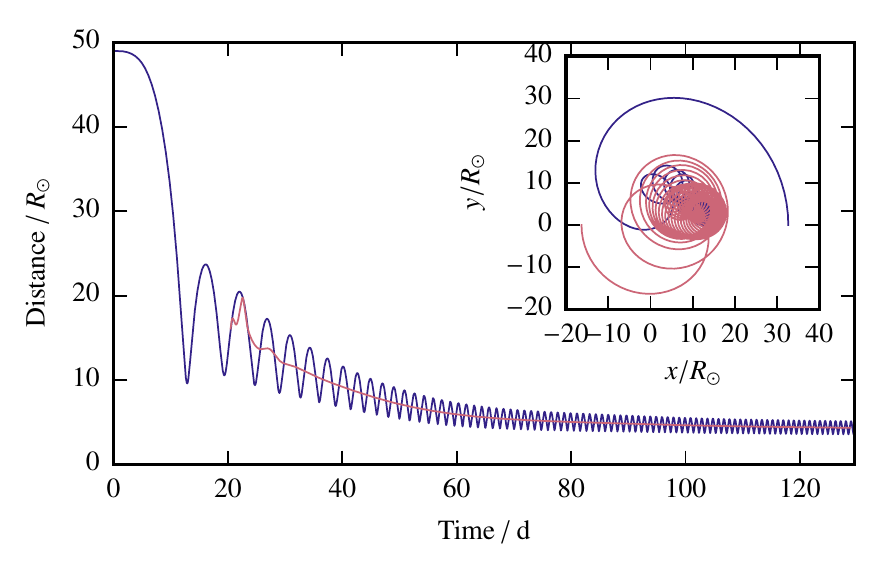}
  \caption{Distance of RG core and companion (blue) and major semi-axis (red)
  in solar radii over time in days. The inset shows the positions of the RG core
  (red) and companion (blue) in the $x$-$y$ plane up to 80~d.}
  \label{fig:distance}
\end{figure}

The simulation starts with tidal deformation of the envelope and mass accretion
on the secondary. When the accretion stream hits itself, shocks emerge and the
orbit shrinks rapidly by a factor of five during the first revolution 
(Fig.~\ref{fig:distance}). This fast spiral-in slows down after a few orbits and
the separation of the RG core and the companion decreases much slower at the end
of the simulation. At this point, the system evolved for over 80
orbits and the separation is about $4.3 R_\odot$, a factor of 10 smaller than in
the beginning. The initially circular orbit becomes eccentric ($e \approx
0.5$), circularizes somewhat in the course of the spiral-in, and the
eccentricity settles at a value of 0.18. Thus, the orbit is rather eccentric
compared to the simulation by \citet{ricker2012a} with $e=0.08$
which may be due to different initial conditions.
At the end of the simulation, the time scale of orbital decay ($-a/\dot{a}$,
$a$: semi-major axis) grows to $\sim 1.5$~yr.

\begin{figure*}[tbp]
  \centering
  \includegraphics[width=\textwidth]{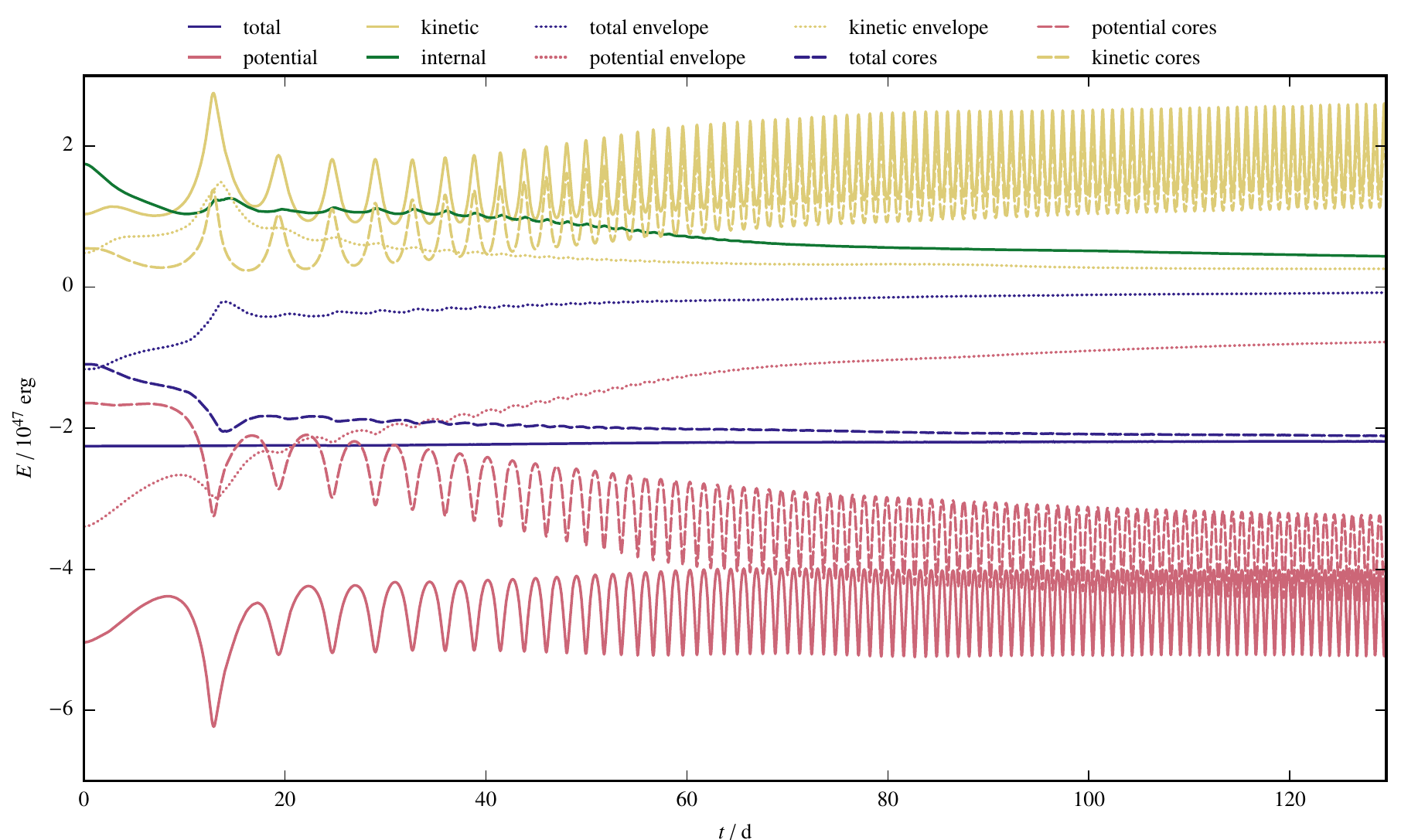}
  \vspace*{-5mm}
  \caption{Energy budget during inspiral. Shown are the total (blue), potential
  (red), kinetic (yellow), and internal energy (green) for the gas of the
  envelope (dotted), the RG core and the companion (dashed), and summed values 
  (full lines). The summed value of the internal energy is the same as for the
  gas since the RG core and the companion are gravitation-only particles. The
  conservation of the total energy is better than 3\% during the simulation.}
  \label{fig:energybudget}
\end{figure*}

The energy budget during the simulation is shown in Fig.~\ref{fig:energybudget}
for the gas cells, the RG core and its companion, and the sum of both. In our
simulation, the core cells interact only gravitationally, hence, they do not
possess internal energy; thus the total internal energy is given by the gas of
the RG envelope only. Although large amounts of potential and kinetic energy are
converted into one another (up to $1.5\times10^{47}$~erg in one orbit, i.e.\ 65\%
of the total energy), the total energy is conserved to better than 3\% during
the whole run. This means that our resolution is sufficient to accurately
represent the
regions where gravity is coupled strongly to the hydrodynamics and where the
conversion between potential and kinetic energies takes place\footnote{This is
not trivial: due to the different discretization of gravity and hydrodynamics,
problems with large conversions of potential energy into kinetic or internal
energy can lead to substantial errors in the total energy, when the resolution
is too low (see the Evrard collapse example in \citealp{springel2010a}).}. 

During the simulation, energy is transferred from the binary system of the RG
core and the companion to the envelope: its binding energy is
raised from $-1.2\times10^{47}$erg in the beginning to $-7.7\times10^{45}$~erg
in the end of the simulation. This energy is mainly taken from the potential
energy of the binary system of the RG core and the companion due to the
shrinking orbit. The internal energy of the envelope decreases by
$1.3\times10^{47}$~erg because of its expansion. Although the total
energy of the envelope is negative at the end of the simulation, $0.1M_\odot$ of
the envelope gets unbound, about 8\% of its mass.  Most of this material is
expelled during the first 40~d. After this, the mass loss rate settles to about
$0.015M_\odot$~yr$^{-1}$. If this mass loss rate is sustained, the envelope may
be ejected in roughly 100~yr.  Similar to the simulations by \citet{passy2012a}
and \citet{ricker2008a,ricker2012a}, only a small fraction of the envelope is
unbound during the fast spiral-in, although more mass is lost in their
simulations at a higher rate.  

Since systems similar to the final system of our simulation are observed (e.g.,
J0755+4800 from \citealp{gianninas2014a}; a $\sim0.4M_\odot$ + $\sim1M_\odot$
system in a $\sim3R_\odot$ orbit), but in a shorter orbit and with the envelope
ejected, additional mechanisms must contribute to the evolution that we do not
capture in our simulation.  This can either be processes acting on longer time
scales (up to the thermal time scale) or additional microphysical effects, such
as recombination (compare the SPH simulations by \citealp{nandez2015a}).

\begin{figure*}[tbp]
  \centering
  \includegraphics[width=\textwidth]{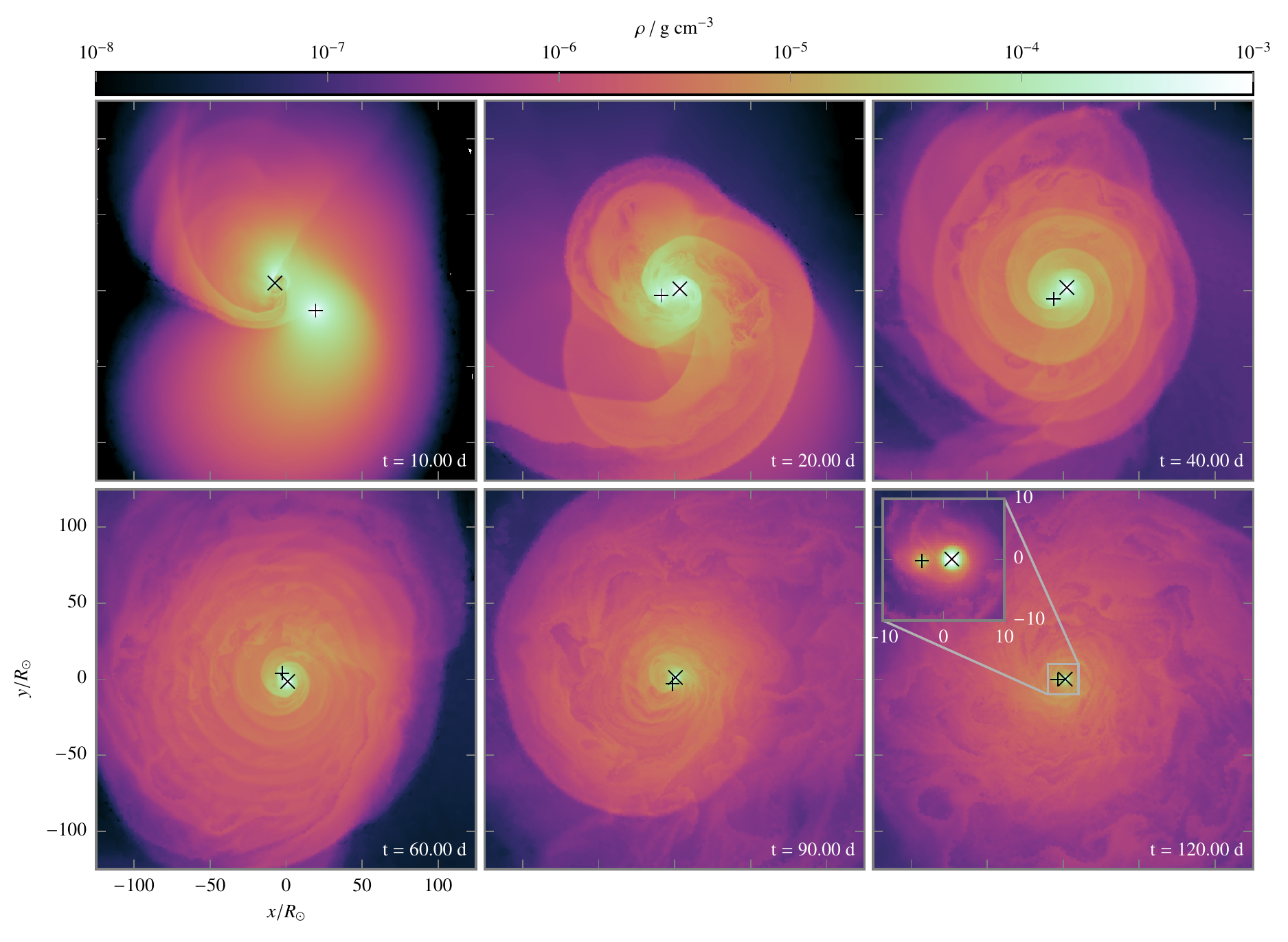}
  \vspace*{-5mm}
  \caption{Time series of density snapshots in the $x$-$y$ plane during spiral-in at
  six different times. The $\times$ marks the position of the companion, the $+$
  marks the position of the RG core. All plots are centered on the center of
  mass of the RG core and the companion. The inset in the last panel shows the
  central region of about $20R_\odot$ with the color scale ranging from
  $10^{-6}$ to $10^{-3}$~g~cm$^{-3}$.}
  \label{fig:timeseries}
\end{figure*}

The dynamics of the spiral-in is illustrated in Fig.~\ref{fig:timeseries} as a
series of density slices in the $x$-$y$ plane. During the first orbit, the
companion plunges into the envelope and an accretion stream onto it
builds up.  After 10~d
(Fig.~\ref{fig:timeseries}, upper left), an accretion shock is visible
that results in a tidal arm moving outwards. Most of the material that is
unbound at the end of the simulation stems from this first interaction.
During the second orbit, the distance
between the RG core and the companion has decreased by a factor of about 5
compared to their initial separation and
the two compact components revolve in an eccentric orbit. The shock created by
the companion reaches the inner part of the envelope while a second shock
is caused by the motion of the RG core. After about 2 orbits (20~d,
Fig.~\ref{fig:timeseries}, upper center), the shock created by the RG core
overtakes the first tidal arm caused by the companion. The density field in the
regions between the shocks does not show distinct features. After almost seven
orbits (40~d, Fig.~\ref{fig:timeseries}, upper right), a layered structure
emerges that is created by spiral shocks continuously driven outwards by both
the RG core and the compact companion. Shear flows between neighbouring shocks
cause Kelvin-Helmholtz instabilities in the outer part of the
spiral structure.  The spiral
structure of the shocks tightens and shear instabilities grow stronger at
60~d (Fig.~\ref{fig:timeseries}, bottom left). 
At later times (90~d, Fig.~\ref{fig:timeseries}, bottom center), shear
instabilities of adjacent layers overlap and form a large-scale instability
that connects several regions of the spiral structure. New shocks are still
created in the inner part by the RG core and the companion. 
Near the end of the simulation (120~d, Fig.~\ref{fig:timeseries}, bottom
right), the spiral shock structure is not visible anymore. Instead, 
large-scale instabilities have emerged and dominate the flow pattern.  The
central part of the domain around the compact
components is still well resolved (see inset of the bottom right panel of
Fig.~\ref{fig:timeseries}). The flow between the RG core and the companion
remains smooth; shocks begin outside the innermost region.  During the
evolution, the flow shows some asymmetries caused by the first tidal arm which
is ejected in the negative $x$ and $y$ directions, resulting in a relocation of the
RG core and the companion in the opposite direction.

\begin{figure*}[ptb]
  \centering
  \includegraphics[width=\textwidth]{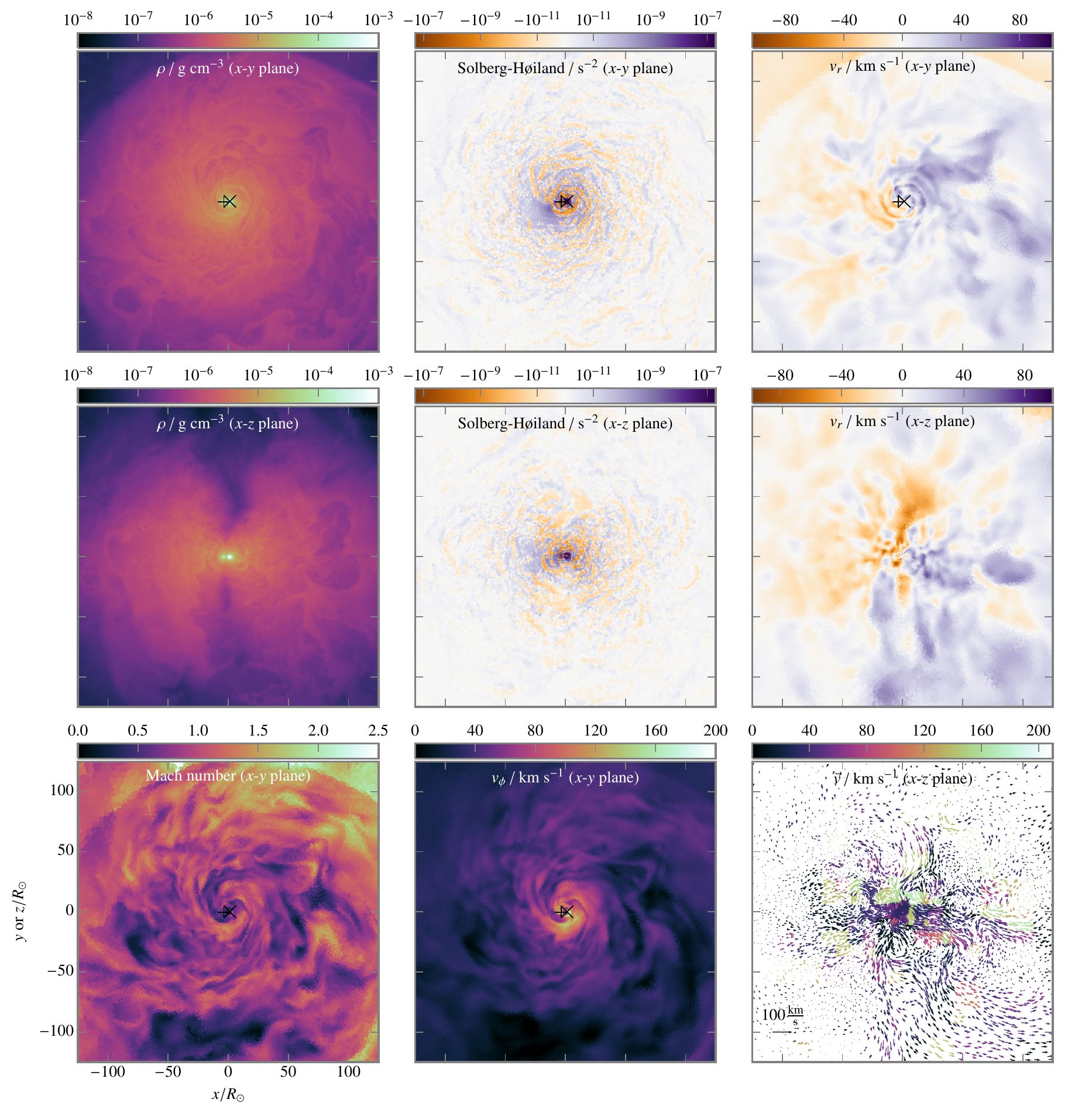}
  \caption{Late snapshot at 120~d. 
    Shown are the density $\rho$ in the $x$-$y$
    (upper left) and $x$-$z$~plane (middle left), the Solberg-H\o{}iland criterion 
    from Eq.~(\ref{eq:solberg}) using a symmetric logarithmic color coding (blue:
    stable; orange: unstable) in the $x$-$y$ (upper center) and $x$-$z$~plane
    (middle center), 
    the radial component of the velocity
    $v_r = \vec{v} \cdot \vec{e}_r$ in the $x$-$y$ (upper right) and $x$-$z$~plane
    (middle right), 
    the Mach number (lower left), 
    the angular component of the velocity
    $v_\phi = \vec{v} \cdot \vec{e}_\phi$ (lower center), and a vector plot of
    the velocity in the $x$-$z$~plane (lower right), with the color coding
    indicating the magnitude of the velocity in km~s$^{-1}$.
    The $\times$ marks the position of
    the companion, the $+$ marks the position of the RG core. All plots are
    centered on the center of mass of the RG core and the companion.}
  \label{fig:final}
\end{figure*}

The structure of the envelope at the end of the simulation after 120~d is
shown in Fig.~\ref{fig:final}. The density slice in the $x$-$y$ plane
(Fig.~\ref{fig:final}, upper left) shows that the layered shock structure is
only retained in the innermost region. In the outer region, it is
washed out and the flow is dominated by large-scale instabilities. 
In the $x$-$z$ plane (see middle left in Fig.~\ref{fig:final}), the outflow is
concentrated mostly around the equatorial plane in a toroidal structure.
Shocks generated in the inner part are washed out by the shear instability in
the outer region.
To assess the convective stability of the envelope, we employ
the Solberg-H\o{}iland criterion \citep[e.g.,][]{kippenhahn2012a}
that predicts convective stability for
\begin{equation}
-\frac{\vec{g}\cdot\nabla s}{c_p} + \frac{1}{\varpi^3}\frac{\partial
j^2}{\partial \varpi} > 0,
  \label{eq:solberg}
\end{equation}
where $\vec{g}$ denotes gravitational acceleration, $s$ entropy, $c_p$ specific
heat capacity at constant pressure, $j$ specific angular momentum, and $\varpi$
distance from rotational axis. This quantity is shown in Fig.~\ref{fig:final}
for the $x$-$y$ plane (upper center) and the $x$-$z$ plane (middle center).
The flow is stabilized by the increase of
specific angular momentum (second term) in a sphere of $\sim 7 R_\odot$ around
the center of mass that is located to the lower left compared to the cores in
Fig.~\ref{fig:final} (upper center). 
Farther away from the center of mass, the specific angular
momentum is nearly constant, and the impact of the second term in the
Solberg-H\o{}iland criterion decreases rapidly. Apart from a small region of
stability ($\sim 3 R_\odot$) around the RG core and the companion, regions of
stability and instability alternate over the envelope because of the
hydrodynamical flows. The situation is similar perpendicular to the orbital
plane, where unstable regions can be found throughout the toroidal structure.
The growth time scale associated to the unstable regions is $\lesssim 100$\,d;
thus, we conclude that large parts of the envelope should be convectively
unstable.
The radial velocity in the orbital plane (Fig.~\ref{fig:final}, upper right)
displays some regions with inflows in the left hemisphere but outflows in the
rest of the envelope.  The inflow is probably caused by the initial plunge-in of the
companion.  The layered structure of the shocks is visible in the inner part as
jumps, but it is overlaid by the instability farther out. 
In the $x$-$z$ plane the radial velocity (Fig.~\ref{fig:final}, middle right)
shows mostly inflows in the upper and left hemispheres and outflows in the lower
and right hemispheres.
This pattern is also seen in the vector plot of the velocity in this plane
(Fig.~\ref{fig:final}, lower right) that additionally shows a complex flow
structure with whirls corresponding to the instabilities in the flow.
This complex structure makes it difficult to predict the further evolution of
energy transport in this plane.
In the region of the developing instability, the flow is mostly
subsonic (Fig.~\ref{fig:final}, lower left), whereas it is transsonic in most
other parts of the envelope and supersonic behind shocks in the outer regions.
The envelope is still mostly co-rotating, as can be
seen in the angular component of the velocity (Fig.~\ref{fig:final}, lower
center), although the velocity is rather small in the region of the
instability. Especially in the inner part, adjacent layers can be found with
differing velocities, resulting in shear flows.

The development of large-scale flow instabilities and an inverse entropy
gradient indicate the onset of turbulent convection in the differentially rotating
envelope. This supports the assumptions of \citet{meyer1979a} of a
co-rotating interior and a differentially rotating envelope with angular
momentum transport mediated by convection in the envelope.

Large-scale flow instabilities have not been observed in hydrodynamics
simulations before. \citet{passy2012a} show a density distribution for their
256$^3$ grid run after about 5 orbits, where only spiral shocks with smooth
regions in between are visible (see their Figure~6).  The density slice of the
AMR simulations by \citet{ricker2012a} after roughly 5 orbits displays features
that may be caused by shear flows between adjacent spiral shocks. Their
simulation was stopped at this instant and it is unclear if large-scale
instabilities would have emerged in the further evolution of the model.  We
suspect, however, that the development of shear instabilities is not seen due to
large numerical diffusion in the SPH simulations and due to the background
velocity field in the grid simulations. The violation of Galilean invariance in
conventional grid-based hydrodynamics codes (when altering the background
velocity at the same resolution) suppresses Kelvin-Helmholtz
instabilities on a static mesh; this is illustrated in Figure~33 of
\citet{springel2010a}. The numerical scheme of \textsc{arepo} is Galilean
invariant, thus, shear instabilities may also develop on top of background
velocities.

\section{Conclusions}
\label{sec:conclusions}

In this \emph{Letter}, we explore the hydrodynamics of the rapid spiral-in
during a CE event using the moving-mesh code \textsc{arepo}.  The combination of
the nearly Lagrangian mesh motion and the Galilean-invariant scheme enables us
to resolve the hydrodynamical structure in unprecedented detail, and complements
recent hydrodynamics simulations \citep{passy2012a,ricker2012a}.  In particular,
we observe, for the first time, the emergence of large-scale flow
instabilities. These are caused by shear between adjacent layers of the shock
spiral that is created by the in-fall of the companion. These instabilities
indicate the onset of turbulent convection, with important consequences for the
further evolution of the system by, e.g., altering the energy transport on
thermal time scales.

In terms of global quantities, we confirm earlier simulations
\citep{ricker2008a,ricker2012a,passy2012a}: only a small fraction of the envelope
mass is ejected on a dynamical timescale and the final separation is larger than
observed.  This may be due to the envelope ejection proceeding on a much longer
time scale than that followed in our simulation.  It is also possible that we
miss other processes driving the loss of the envelope, such as recombination.

As a next step, we will improve the modeling of additional microphysical
effects, including recombination, and examine different parameters (orbital
parameters, masses) to link the final system characteristics to them in a
systematic way.  This opens up the exciting prospect to directly connect the
hydrodynamics simulations to binary stellar evolution.

\acknowledgments{%
  We thank Jean-Claude Passy, Orsola de Marco, and Philipp Edelmann
  for helpful discussions.
  STO acknowledges support from the Studienstiftung des deutschen Volkes.
  FKR acknowledges support by the DAAD/Go8 German-Australian exchange program,
  and by the ARCHES prize of the German Ministry of Education and Research
  (BMBF).
  RP and VS acknowledge support  by the European Research Council
  under ERC-StG grant EXAGAL-308037 and by the Klaus Tschira
  Foundation.
  This work was also supported by  the graduate school ``Theoretical
  Astrophysics and Particle Physics'' at the University of W\"urzburg (GRK
  1147).
  For data processing and plotting, we used NumPy and SciPy
  \citep{oliphant2007a}, IPython \citep{perez2007a}, and Matplotlib
  \citep{hunter2007a}.
}

\bibliographystyle{apj}

\end{document}